# Anomalously Enhanced Diffusivity of Moiré Excitons via Manipulating the Interplay with Correlated Electrons

Li Yan[1,2#], Lei Ma[1,2#], Yuze Meng[1#], Chengxin Xiao[3#], Bo Chen[1], Qiran Wu[4,5], Jingyuan Cui[4,5], Qingrui Cao[1], Rounak Banerjee[6], Takashi Taniguchi[7], Kenji Watanabe[8], Seth Ariel Tongay[6], Benjamin Hunt[1], Yong-Tao Cui[4,5], Wang Yao[3*], Su-Fei Shi[1*]

1. Department of Physics, Carnegie Mellon University, Pittsburgh, PA, 15213, USA
2. Department of Chemical and Biological Engineering, Rensselaer Polytechnic Institute, Troy, NY 12180, USA
3. Department of Physics, University of Hong Kong, Hong Kong, China
4. Department of Physics and Astronomy, University of California, Riverside, California, 92521, USA
5. Department of Materials Science and Engineering, University of California, Riverside, California, 92521, USA
6. School for Engineering of Matter, Transport and Energy, Arizona State University, Tempe, AZ 85287, USA
7. International Center for Materials Nanoarchitectonics, National Institute for Materials Science, 1-1 Namiki, Tsukuba 305-0044, Japan
8. Research Center for Functional Materials, National Institute for Materials Science, 1-1 Namiki, Tsukuba 305-0044, Japan

[#] These authors contributed equally to this work
[*] Corresponding authors: sufeis@andrew.cmu.edu; wangyao@hku.hk

## Abstract

**Semiconducting transitional metal dichalcogenides (TMDCs) moiré superlattice provides an exciting platform for manipulating excitons. The in-situ control of moiré potential confined exciton would usher in unprecedented functions of excitonic devices but remains challenging. Meanwhile, as a dipolar composite boson, interlayer exciton in the type-II aligned TMDC moiré superlattice strongly interacts with fermionic charge carriers. Here, we demonstrate active manipulation of the exciton diffusivity by tuning their interplay with correlated carriers in moiré potentials. At fractional fillings where carriers are known to form generalized Wigner crystals, we observed suppressed diffusivity of exciton. In contrast, in Fermi liquid states where carriers dynamically populate all moiré traps, the repulsive carrier-exciton interaction can effectively reduce the moiré potential confinement seen by the exciton, leading to enhanced diffusivity with the increase of the carrier density. Notably, the exciton diffusivity is enhanced by orders of magnitude near the Mott insulator state, and the enhancement is much more pronounced for the 0-degree than the 60-degree aligned $WS_2/WSe_2$ heterobilayer due to the more localized nature of interlayer excitons. Our study inspires further engineering and controlling exotic excitonic states in TMDC moiré superlattices for fascinating quantum phenomena and novel excitonic devices.**

**Main Text**

Transitional metal dichalcogenide (TMDC) moiré superlattice ushers in unprecedented opportunities not only for engineering electronic structures[1,2,3,4,5] but also for enabling new excitonic states[6,7,8,9,10,11,12,13,14,15], in which a periodic potential landscape with desired periodicity and potential profile can be designed to control exciton motion and even form exciton superlattices. However, active control of the exciton motion, critical for eventual device applications, often requires in-situ tunability, such as the tunable twist angle[16,17], which is challenging to achieve.

Meanwhile, it has been shown that various TMDC moiré superlattices can enable flat moiré bands[1,18,19,20,21,22,23,24], in which the electron kinetic energy is significantly suppressed, and the strong Coulomb interaction is enhanced due to the reduced screening in two-dimensional (2D) materials. The resulting enhanced electron correlation, characterized by the ratio of Coulomb interaction to kinetic energy, leads to exciting opportunities for realizing and exploring quantum-correlated states, such as Mott insulators[19,25], Wigner crystals[18,20,21], and excitonic insulators[26,27,28].

The interlayer excitons in a type-II aligned TMDC moiré heterojunction, with electron and hole separated in different layers[9,29], inherit this correlation and strongly interact with electrons[7]. This strong interaction is manifested in the significant blueshifts of interlayer exciton PL at the correlated electronic states[7,30]. This inspires us to employ the electron-exciton interaction to control the exciton dynamics such as exciton diffusion[31,32,33,34,35,36,37,38,39,40,41] in the angle-aligned $WS_2/WSe_2$ moiré heterojunction, by directly measuring the interlayer exciton diffusion length and extracting the diffusivity as a function of the electrostatic doping. We found that when electrons crystalize at fractional fillings, forming a generalized Wigner crystal, the excitons scatter off electrons effectively interact with the electron crystal as a whole, which results in increased backscattering and limits the mobility of the exciton, leading to suppressed exciton diffusivity. In stark contrast, in Fermi liquid states where the electrons dynamically populate all moiré traps, the electron-exciton repulsion effectively reduces the moiré potential confinement of excitons and mobilizes exciton, leading to enhanced diffusivity of exciton. Strikingly, the exciton diffusivity can be enhanced by as high as more than one order of magnitude when the electron doping density is less than one per moiré cell but close to the Mott insulator transition, approaching about 4 times enhancement for the 60-degree (H-stacked) aligned $WS_2/WSe_2$ device and about 50 times for the 0-degree (R-stacked) device. The more significant enhancement in diffusivity in R-stacked $WS_2/WSe_2$ moiré heterojunctions arises from the different nature of the two different stacked configurations: the interlayer exciton is more extended in the H-stacked $WS_2/WSe_2$ as the associated electron and hole are at different moiré sites[7,30]. The diffusivity can be enhanced even more, reaching an enhancement as large as 150-fold compared with that at charge neutral scenario, when the doping exceeds one electron per moiré cell, where the interlayer exciton goes to the excited state due to the strong electron-exciton onsite repulsion. The drastic modulation of the moiré exciton diffusivity by tuning electrostatic doping and correlated electrons ushers in an exciting route of engineering exciton dynamics via quantum many-body interactions[42] and can be explored for new quantum states and novel optoelectronic devices.

**Exciton diffusion at the integer and fractional fillings**

A typical dual-gated $WS_2/WSe_2$ moiré heterojunction device is schematically shown in Fig. 1a. The doping-dependent PL spectra of the interlayer exciton in $WS_2/WSe_2$ moiré heterojunction have been extensively studied[6,7,30,43,44]. In Fig. 1, we excite a dual-gate H-stacked device (H1) with a continuous wave (CW) laser with the energy centered at 1.699 eV, which is in resonance with the ground state of moiré intralayer excitons in the angle-aligned $WS_2/WSe_2$ heterojunction[18].

The diffusion of excitons has been studied with various methods[32,33,36,40,41,45]. Here, we optically excited excitons (schematically shown in Fig. 1b) and measure the broadened spatial PL image, which we then compare with the size of the laser spot[34,35,36,37,41]. For the excitation power of 0.1 µW, which is well within the linear response regime to avoid nonlinear exciton dynamics (see Supplementary Section 3 for the power dependence of the PL), we found a broadened PL image spot of (~ 2.5 µm) compared with that of the laser spot of ~ 1.5 µm (Fig. 1b). In this linear regime, the broadened PL profile with a Gaussian width σ (Supplementary Section 6) compared with the laser beam width ($σ_0$) can be well described by the diffusion of interlayer excitons, with the diffusion length $l_d$ expressed as $l_d^2 = σ^2 − σ_0^2 = 2Dτ$ (1), where D is the diffusion constant (diffusivity), and τ is the lifetime of the exciton. We further study the diffusion of interlayer excitons at different doping by performing spectral-resolved diffusion measurements of interlayer excitons. Fig. 1c-f shows spectral-resolved diffusion images with different filling factors, which normalizes the doping with the moiré superlattice density, with n=1(-1) being one electron (hole) per moiré cell. It is evident that the diffusion length increases when the doping is increased from n=0 (Fig. 1f) to n=1 (Fig. 1c). However, one can also notice that the exciton diffusion length is suppressed at n=1/3 (Fig. 1d), which corresponds to the previously established generalized Wigner crystal insulator[7,18,21], with one electron shared by three moiré unit cells. At n=1, a new exciton resonance at a higher energy state (1.47 eV) appears (Fig. 1c), which is due to the formation of an electronic Mott insulator at the doping density of one electron per moiré unit cell [cite our nature physics paper]. The doping-modulated exciton diffusion length change can also be seen in Fig. 1g, in which the spatial distribution of the PL (line cuts from Fig. 1c-f) is plotted along with the spatial distribution of the laser.

Figs. 2a,b show the doping-dependent interlayer exciton PL spectra from one H-stacked device (H1) and one R-stacked device (R1), respectively. The correlated insulating states at integer fillings (n=±1, ±2) and fractional fillings (n=±1/3, n=±2/3, and more) can be clearly seen in Figs. 2a,b, which manifests the high quality of the moiré bilayer devices. We then study the spatially resolved PL of the interlayer exciton at different doping, and we extract the diffusion length ($l_d^2 = σ^2 − σ_0^2$) and plot it as a function of doping in Figs. 2c,e, respectively, with the zoom-ins (Figs. 2d,f) amplifies the fine features between the filling factor -1 and 1. The abrupt blueshifts of the interlayer exciton PL are a result of onsite repulsion between the correlated electron and exciton, as we reported previously[7,43]. The resulting interlayer excitons are at excited states, and we label them in different colors to manifest their different nature: black for -1<n<1, blue for -2<n<-1 and 1<n<2, and magenta for n>2 and n<-2.

The most pronounced feature of the H-stacked device (H1, Fig. 2c,d) is the abrupt decrease of diffusion length at n= ±1, where the low energy peak (-1<n<+1, red dots) transits into the high energy peak (n<-1 or n>1, blue dots). For the R-stacked device (R1, Fig. 2e,f), this drop in the diffusion length is less drastic, and the blueshift of PL peak energy is less pronounced.

We also find the suppression of the exciton diffusion length for both H- and R-stacked devices (red and blue dots in Figs. 2d,f, respectively) closely aligns with these fractional filling states, showing that the doping dependence of interlayer exciton diffusion length can resolve correlated insulating states, similar to that the gate-dependent PL spectra (Figs. 2a,b). This observation has been found in more than 22 angle-aligned $WS_2$/$WSe_2$ devices that we have measured, including 14 H-stacked and 8 R-stacked devices. We further fabricated a single-gated device (H2), which we can measure local conductance by microwave impendence microscopy (MIM). The measured diffusion length suppression also correlates well with the fractional filling states obtained with MIM measurements (see Supplementary Section 2 for details).

**Lifetime and diffusivity**

To investigate these features, we perform time-resolved PL (TRPL) measurements for H1 (Fig. 3a) and R1 (Fig. 3c), respectively. The lifetimes of the exciton ($\tau$) for both devices drop abruptly at n=±1 (Fig. 3a,c) due to the excited state nature of the interlayer exciton in the presence of strong onsite electron-exciton repulsion[7,44]. Based on Eqn. (1), The abrupt decrease in lifetime at n=±1 contributes to the sudden drop of exciton diffusion length at the Mott insulator states. We note a much more dramatic drop at n=±1 for the H-stacked device (H1, Fig. 3a) than the R-stacked device (R1, Fig. 3c), as the lifetime decrease is more significant for the device H1 compared to the R1.

However, the diffusion length dips observed at fractional fillings for both R and H stack samples cannot be attributed to the lifetime change, as the lifetimes at the fractional fillings (shown in Figs. 3a,c) increase, likely due to the suppressing of nonradiative channels once the correlated insulators form[46]. From Eqn. (1), such decrease in diffusion length can only be explained by the reduced diffusivity at the fractional fillings. If we combine the diffusion length obtained in Fig. 2 and extrapolate the lifetime data from Figs. 3a,c, we can obtain the exciton diffusivity following Eqn. (1) as $D = (\sigma^2 - \sigma_0^2)/2\tau$. We plot the diffusivity as a function of the doping for device H1 (Fig. 3b) and R1 (Fig. 3d), respectively.

**Enhancement and suppression of diffusivity for different doping**

The diffusivity in both devices is evidently suppressed at n=0, which we attribute to the moiré potential confinement impeding exciton diffusion, consistent with previous reports[31,40,41,47]. Besides the diffusivity suppression at the fractional filling states, Fig. 3b,d show that the diffusivity increases by around 4 times for device H1 and around 50 times for R1 when the doping is increased from n=0 to close to n= ±1, which can be better seen from the insets of Fig. 3b,d (red dots). To understand this behavior, we analyze the effective potential experienced by the interlayer exciton upon increasing electrostatic doping. Take electron doping as an example: by introducing electrons into the system, the strong repulsion between electrons and interlayer excitons effectively lowers the moiré potential confinement of interlayer excitons. It renders them easier to diffuse, as schematically shown in Figs. 3e,f, in which the effective potential well depth is lowered for excitons at n=0.2 than that at n=0, and the diffusivity is increased (black arrows in Figs. 3e,f b).

It is interesting to note that the enhancement factor near the Mott insulator state at n=±1 is drastically different for devices H1 and R1, and this difference has been reproduced in another H-stacked device (H2) and R-stacked device (R2) (Supplementary Section 8). This striking difference is likely attributed to the more extended nature of interlayer excitons in the H-stacked devices. The Coulomb repulsion between electron and exciton that lowers the effective moiré potential depth is a sensitive function of screening, which is increased at increased doping. However, the interlayer excitons in the R-stacked moiré structure are more localized and less affected by the screening, which can be seen from less sensitivity of PL resonances to the doping (Figs. 2a,c) as reported earlier[7,30]. As a result, the Coulomb repulsion remains strong for the R-stacked device even at the increased electron doping, leading to much increased exciton diffusivity. This finding suggests that the R-stacked device is more promising for excitonic devices that require drastic modulation of exciton diffusivity.

The suppressed diffusivity by the correlated electrons at fractional fillings (the insets in Fig. 3b,d) can also be explained by the abovementioned electron-exciton interaction. At the fractional fillings such as n=1/3, electrons form periodic crystals with electrons localized in one-third of the moiré cells. This is different from the free electron doping case, in which the repulsive interaction is from delocalized electron, which averaged to be 1/3 electron per moiré cell. Additionally, as the electrons form the periodic generalized Wigner crystals, excitons scattering off electrons have to interact with the electron crystal as a whole, which possesses large mass inertial. The resulting back-scattering will further suppress the diffusivity. The electron Fermi liquid to electron crystal transition, therefore, would drive the enhanced diffusivity to suppressed diffusivity transition of the interlayer excitons due to the modification of the effective moiré potential owing to the unique electron-exciton interaction in this system. Our Morte Carlo simulations show the suppression of the exciton diffusivity at the correlated insulator states, as shown in

Supplementary Fig. 7. Detailed modeling and calculations can be found in Supplementary Section 7.

An even more striking enhancement of exciton diffusivity can be found as the doping increases beyond the Mott insulator states (n>1 or n<-1). The diffusivity exhibits around two orders of magnitude increase from n=0 to n>1 (62 times for the H1 devive and 200 times for R1). Such a drastic increase can be observed in all the samples we have measured (see Supplementary Section 8) and is consistent with the excited state nature of the exciton for n>1 or n<-1, where the significantly blue-shifted exciton PL energy suggested that the interlayer exciton is at an excited state, with a strong onsite repulsion from the correlated electrons forming a fermionic Mott insulator. The blue shifted exciton energy, as large as 20-40 meV, suggests that the excitons are at much-elevated energy and are less confined by the moiré potential, hence the much-enhanced diffusivity. The diffusivity is even more enhanced for the exciton with the resonance energy further blue-shifted at n>2 or n<-2, confirming this interpretation.

**Conclusions**

In summary, our study reveals an intriguing moiré system in which the exciton dynamics can be sensitively tuned through electrostatic doping, owing to the unique exciton-electron interaction stemming from strong electron correlation. The electrostatic doping-enhanced diffusivity is also sensitive to the stacking order of the $WS_2$/$WSe_2$ moiré superlattice, which hosts different interlayer excitons with the layer-separated electron and hole either occupying the same (R-stacked device) or different (H-stacked device) moiré sites. Our work provides critical insight into further engineering exciton dynamics. The active control of interlayer exciton diffusion via electric means is also promising for further realization and implementation of moiré excitonic devices.

## Method

### Device fabrication

We used a dry pick-up method that wsa reported in our previous work[48] to fabricate angle-aligned WS$_2$/WSe$_2$ heterostructures. Gold electrodes are pre-patterned on the Si/SiO$_2$ substrate. The monolayer TMDC, h-BN, and a few-layer graphite flakes were directly exfoliated on silicon chips with 285 nm thermal oxide and then picked up by a polycarbonate (PC)/ polydimethylsiloxane (PDMS) stamp. The fabricated devices were then dried with nitrogen gas and annealed in a vacuum (< 10$^{-6}$ torr) at 250 °C for 8 hours.

### Optical Characterization

For all the optical measurements, the samples were measured in a cryogen-free optical cryostat (Attocube 1000) under 3.6 K for all measurements. A home-built confocal imaging system was used to focus laser onto the sample (with a beam spot diameter ~ 1.5 μm) and collect the optical signal into a spectrometer (Princeton Instruments). The PL and diffusion measurements were performed with a CW laser with the photon energy centered at 1.699 eV. For the diffusion measurements, an achromatic lens with a focal length of 500 mm was used to obtain a 200x magnification. The TRPL measurements were performed with a supercontinuum laser (YSL Photonics) with a 1 MHz repetition rate, and a band pass filter (10 nm window) was used to select the optical excitation centered at 650 nm. The signal was collected by a single-photon detector (PicoQuant) and a picosecond event timer (PicoHarp 300).

### Data Availability

All data that support the plots within this paper and other findings of this study are available from the corresponding authors upon reasonable request.

### Acknowledgments

We thank Chenhao Jin for the helpful discussions. S.-F.S. also acknowledges the support from NSF (Career Grant DMR-1945420, DMR-2104902, and ECCS-2139692). J.C., Q.W., and Y.-T.C. acknowledge support from NSF under awards DMR-2104805 and DMR-2145735. K.W. and T.T. acknowledge support from the Elemental Strategy Initiative conducted by the MEXT, Japan, Grant Numbers JPMXP0112101001 and JSPS KAKENHI, Grant Numbers 19H05790 and JP20H00354. S.T acknowledges primary support from DOE-SC0020653 (materials synthesis), Applied Materials Inc., DMR 2111812, DMR 2206987, and CMMI 2129412

### Author contributions

S.-F. S. conceived the project. Y. M. fabricated heterostructure devices. L. Y. and L.M. performed the optical spectroscopy measurements. Q.W., J. C., and Y.-T. C. performed the MIM measurements. M.B. and S.T. grew the TMDC crystals. T.T. and K.W. grew the BN crystals. Q. C. and B. H. helped with device fabrication. C.X. and Y.W. contributed to the theoretical understanding and simulations. S.-F. S, Y. W., L. Y., L. M, Y. M, and C. X.

analyzed the data. S.-F. S. wrote the manuscript with the help of L. Y., L. M., Y. M., B. C. and input from all authors.

These authors contributed equally: Li Yan, Lei Ma, Yuze Meng, Chengxin Xiao.

**Competing interests**

The authors declare no competing interest.

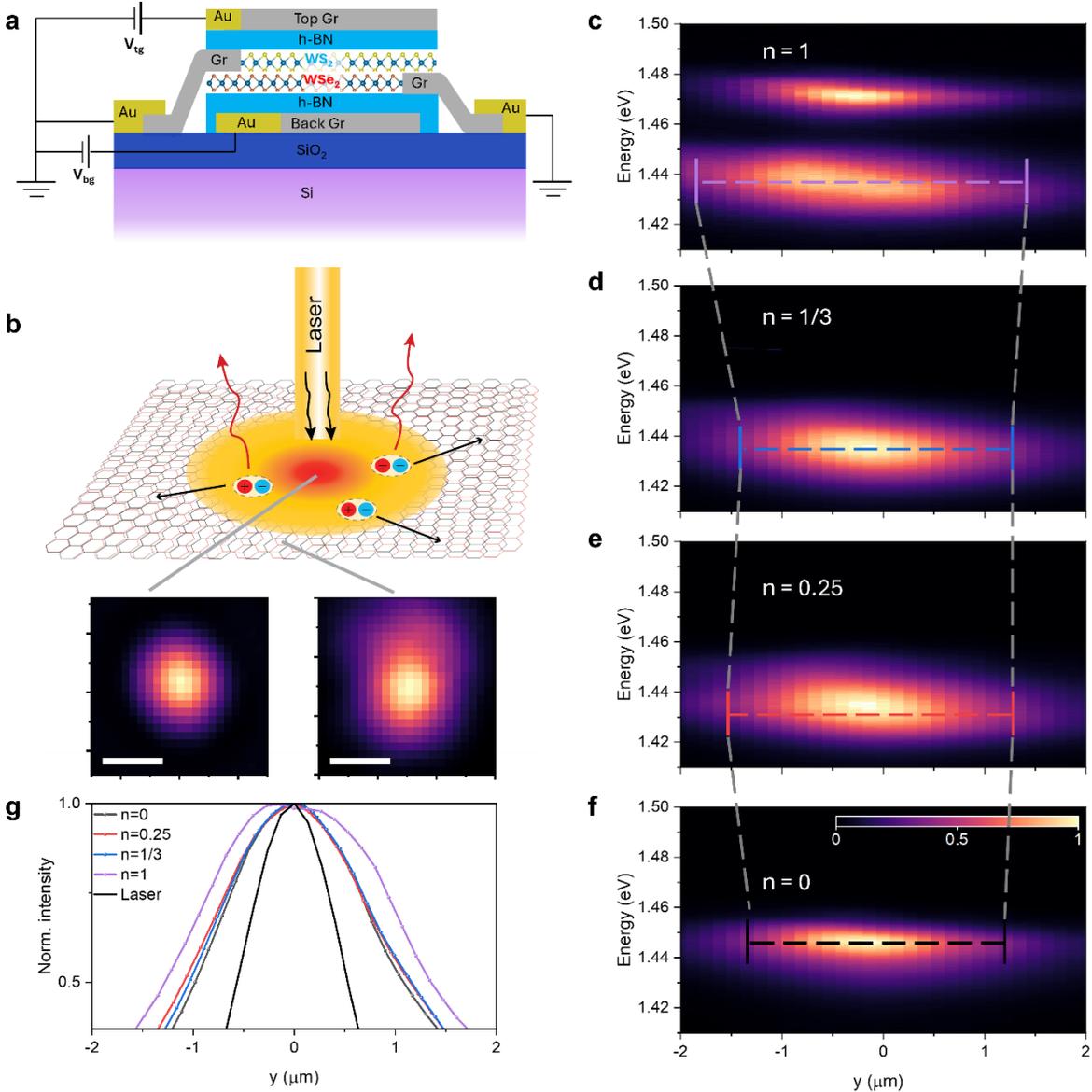

**Figure 1. Interlayer exciton diffusion in a WS$_2$/WSe$_2$ moiré superlattice with tunable doping.** (a) Schematic of a dual-gated WS$_2$/WSe$_2$ moiré superlattice device. (b) Schematic of the exciton diffusion measurement, along with the spatial distribution of laser spot and PL spot. Scale bar: 2 µm. (c-f) Spatially resolved PL spectra of the interlayer exciton at different doping, with the x-axis showing the spatial diffusion of the interlayer exciton. The colored dashed lines overlay on the images show the spatial diffusion by tracing 1/e of the maximum intensity. (g) The radial line-cut of the laser (black) and PL (colored dot) spatial profile for a dual-gated H-stacked device (H1) at different doping. The excitation photon energy is centered at 1.699 eV and the excitation power is 0.1 µW. All data were taken at a temperature of 3.6 K.

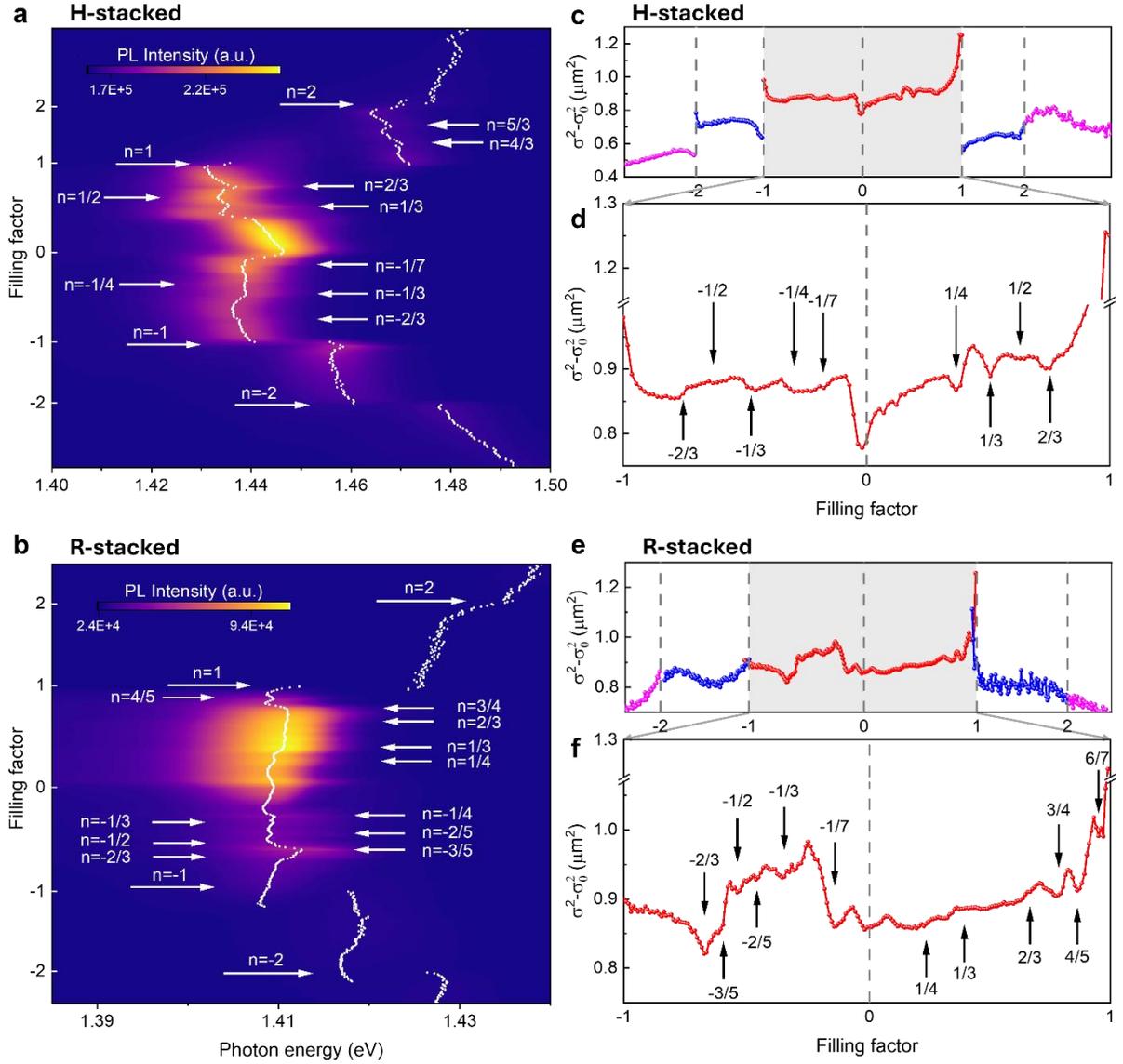

**Figure 2. Interlayer exciton diffusion as a function of doping for H-stacked and R-stacked WS$_2$/WSe$_2$ devices.** (a) and (b) are PL spectra as a function of the filling factor for an H-stacked device (H1) and R-stacked device (R1), respectively. The dotted white lines in the color plots are the extracted PL peak energies. (c) and (e) are the extracted diffusion length for different interlayer exciton species (colored dots) as a function of the filling factor for H1 and R1, respectively. (d) and (f) are the zoom-ins of the gray area of (c) and (e) with black arrows labeling fractional fillings. The excitation power is 0.1 µW for (a), (c), (d) and 0.25 µW for (b), (e), (f).

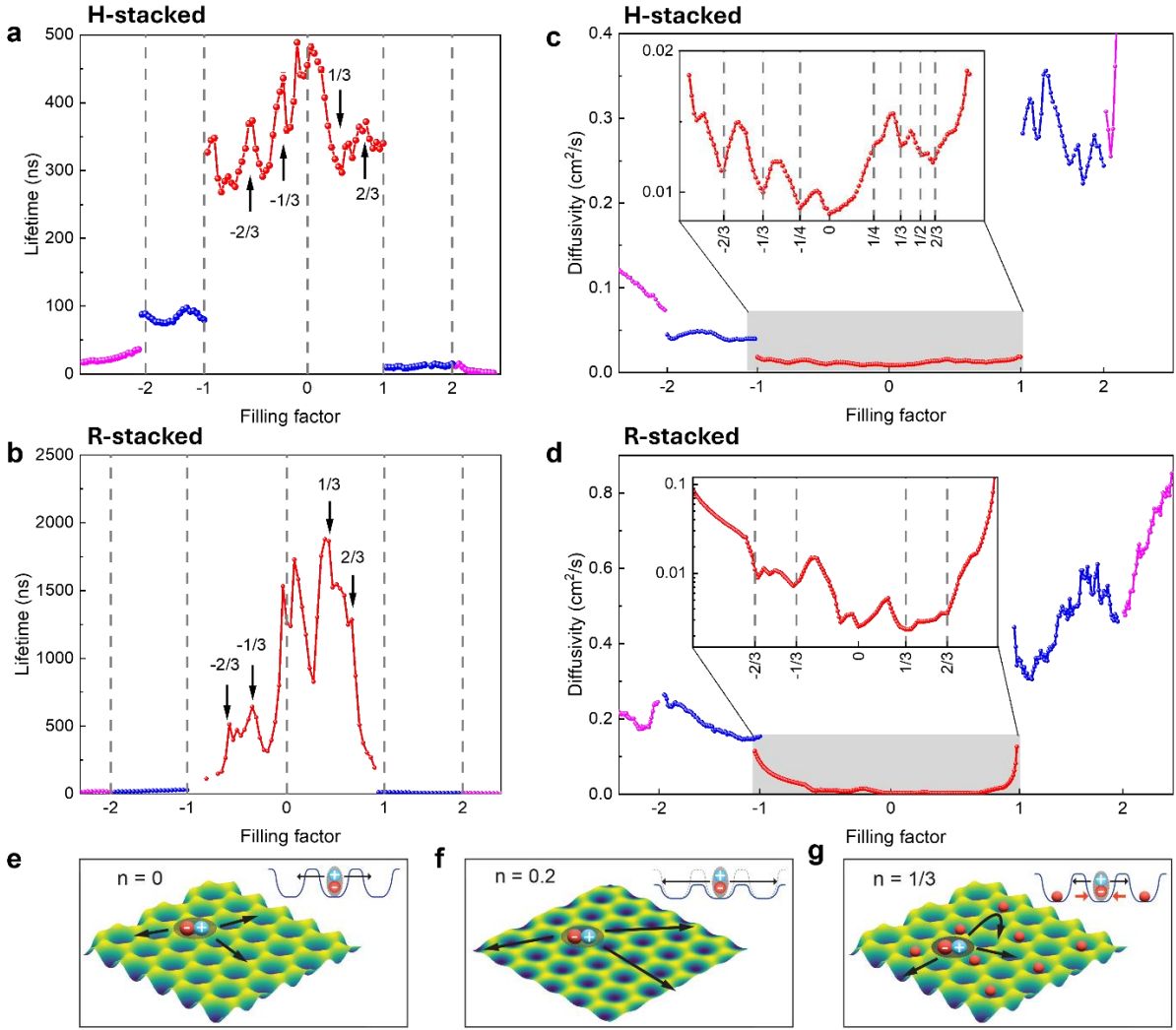

**Figure 3. Lifetime and diffusivity as a function of doping for H-stacked and R-stacked WS$_2$/WSe$_2$ devices.** (a) and (b) are lifetime extracted from TRPL measurements as a function of the filling factor for H1 and R1, respectively. (c) and (d) are diffusivity as a function of the filling factor for H1 and R1, respectively. The insets in (b) and (d) are the zoom-ins of the diffusivity at the filling factor between -1 and 1. (e)-(g) are the schematics of the exciton diffusion at different electron doping.

**References**

1. Zhang, Z. *et al.* Flat bands in twisted bilayer transition metal dichalcogenides. *Nat Phys* ,2020, **16,** 1093–1096

2. Wu, F. *et al.* Hubbard Model Physics in Transition Metal Dichalcogenide Moiré Bands. *Phys Rev Lett* ,2018, **121,**

3. Wu, F. *et al.* Topological Insulators in Twisted Transition Metal Dichalcogenide Homobilayers. *Phys Rev Lett* ,2019, **122,**

4. Devakul, T. *et al.* Magic in twisted transition metal dichalcogenide bilayers. *Nat Commun* ,2021, **12,**

5. Yu, H. *et al.* Giant magnetic field from moiré induced Berry phase in homobilayer semiconductors. *Natl Sci Rev* ,2020, **7,** 12–20

6. Xiong, R. *et al. Correlated insulator of excitons in WSe 2 /WS 2 moiré superlattices. Science* ,2023, **380,**

7. Lian, Z. *et al.* Valley-polarized excitonic Mott insulator in WS2/WSe2 moiré superlattice. *Nat Phys* ,2024, **20,** 34–39

8. Yu, H. *et al. P H Y S I C S Moiré excitons: From programmable quantum emitter arrays to spin-orbit-coupled artificial lattices.* ,2017,

9. Rivera, P. *et al.* Interlayer valley excitons in heterobilayers of transition metal dichalcogenides. *Nature Nanotechnology* ,2018, **13,** 1004–1015

10. Seyler, K. L. *et al.* Signatures of moiré-trapped valley excitons in MoSe2/WSe2 heterobilayers. *Nature* ,2019, **567,** 66–70

11. Tran, K. *et al.* Evidence for moiré excitons in van der Waals heterostructures. *Nature* ,2019, **567,** 71–75

12. Jin, C. *et al.* Observation of moiré excitons in WSe2/WS2 heterostructure superlattices. *Nature* ,2019, **567,** 76–80

13. Alexeev, E. M. *et al.* Resonantly hybridized excitons in moiré superlattices in van der Waals heterostructures. *Nature* ,2019, **567,** 81–86

14. Brotons-Gisbert, M. *et al.* Spin–layer locking of interlayer excitons trapped in moiré potentials. *Nat Mater* ,2020, **19,** 630–636

15. Li, W. *et al.* Dipolar interactions between localized interlayer excitons in van der Waals heterostructures. *Nat Mater* ,2020, **19,** 624–629

16. Tang, H. *et al.* On-chip multi-degree-of-freedom control of two-dimensional materials. *Nature* ,2024, **632,** 1038–1044

17. Kapfer, M. *et al. Programming twist angle and strain profiles in 2D materials*.

18. Regan, E. C. *et al.* Mott and generalized Wigner crystal states in WSe2/WS2 moiré superlattices. *Nature* ,2020, **579,** 359–363

19. Tang, Y. *et al.* Simulation of Hubbard model physics in WSe2/WS2 moiré superlattices. *Nature* ,2020, **579,** 353–358

20. Huang, X. *et al.* Correlated insulating states at fractional fillings of the WS2/WSe2 moiré lattice. *Nat Phys* ,2021, **17,** 715–719

21. Xu, Y. *et al.* Correlated insulating states at fractional fillings of moiré superlattices. *Nature* ,2020, **587,** 214–218

22. Shimazaki, Y. *et al.* Strongly correlated electrons and hybrid excitons in a moiré heterostructure. *Nature* ,2020, **580,** 472–477

23. Li, T. *et al.* Charge-order-enhanced capacitance in semiconductor moiré superlattices. *Nat Nanotechnol* ,2021, **16,** 1068–1072

24. Jin, C. *et al.* Stripe phases in WSe2/WS2 moiré superlattices. *Nat Mater* ,2021, **20,** 940–944

25. Li, T. *et al.* Continuous Mott transition in semiconductor moiré superlattices. *Nature* ,2021, **597,** 350–354

26. Chen, D. *et al.* Excitonic insulator in a heterojunction moiré superlattice. *Nat Phys* ,2022, **18,** 1171–1176

27. Zhang, Z. *et al.* Correlated interlayer exciton insulator in heterostructures of monolayer WSe2 and moiré WS2/WSe2. *Nat Phys* ,2022, **18,** 1214–1220

28. Gu, J. *et al.* Dipolar excitonic insulator in a moiré lattice. *Nat Phys* ,2022, **18,** 395–400

29. Jin, C. *et al.* Ultrafast dynamics in van der Waals heterostructures. *Nature Nanotechnology* ,2018, **13,** 994–1003

30. Wang, X. *et al.* Intercell moiré exciton complexes in electron lattices. *Nat Mater* ,2023, **22,** 599–604

31. Li, Z. *et al.* Interlayer Exciton Transport in MoSe2/WSe2Heterostructures. *ACS Nano* ,2021, **15,** 1539–1547

32. Rossi, A. *et al.* Anomalous Interlayer Exciton Diffusion in WS2/WSe2 Moiré Heterostructure. *ACS Nano* ,2024, **18,** 18202–18210

33. Sun, Z. *et al.* Excitonic transport driven by repulsive dipolar interaction in a van der Waals heterostructure. *Nat Photonics* ,2022, **16,** 79–85

34. Tagarelli, F. *et al.* Electrical control of hybrid exciton transport in a van der Waals heterostructure. *Nat Photonics*, 2023, **17,** 615–621

35. Jauregui, L. A. *et al. Electrical control of interlayer exciton dynamics in atomically thin heterostructures. Science*, 2019, **366,**

36. Wang, J. *et al.* Diffusivity Reveals Three Distinct Phases of Interlayer Excitons in MoSe2/WSe2 Heterobilayers. *Phys Rev Lett*, 2021, **126,**

37. Bai, Y. *et al.* Evidence for Exciton Crystals in a 2D Semiconductor Heterotrilayer. *Nano Lett*, 2023, **23,** 11621–11629

38. Wietek, E. *et al.* Nonlinear and Negative Effective Diffusivity of Interlayer Excitons in Moiré-Free Heterobilayers. *Phys Rev Lett*, 2024, **132,**

39. Wagner, K. *et al.* Diffusion of Excitons in a Two-Dimensional Fermi Sea of Free Charges. *Nano Lett*, 2023, **23,** 4708–4715

40. Yuan, L. *et al.* Twist-angle-dependent interlayer exciton diffusion in WS2–WSe2 heterobilayers. *Nat Mater*, 2020, **19,** 617–623

41. Choi, J. *et al. Moiré potential impedes interlayer exciton diffusion in van der Waals heterostructures. Sci. Adv*, 2020, **6,**

42. Brem, S. & Malic, E. Bosonic Delocalization of Dipolar Moiré Excitons. *Nano Lett*, 2023, **23,** 4627–4633

43. Miao, S. *et al.* Strong interaction between interlayer excitons and correlated electrons in WSe2/WS2 moiré superlattice. *Nat Commun*, 2021, **12,**

44. Park, H. *et al.* Dipole ladders with large Hubbard interaction in a moiré exciton lattice. *Nat Phys*, 2023, **19,** 1286–1292

45. Kulig, M. *et al.* Exciton Diffusion and Halo Effects in Monolayer Semiconductors. *Phys Rev Lett*, 2018, **120,**

46. Tan, Q. *et al.* Layer-dependent correlated phases in WSe2/MoS2 moiré superlattice. *Nat Mater*, 2023, **22,** 605–611

47. Knorr, W. *et al.* Exciton transport in a moiré potential: From hopping to dispersive regime. *Phys Rev Mater*, 2022, **6,**

48. Lian, Z. *et al.* Quadrupolar excitons and hybridized interlayer Mott insulator in a trilayer moiré superlattice. *Nat Commun*, 2023, **14,**